\documentclass[aps,prd,twocolumn,nofootinbib,preprintnumbers,bibnotes,superscriptaddress]{revtex4-2}

\usepackage{graphicx}% Include figure files
\usepackage{dcolumn}% Align table columns on decimal point
\usepackage{bm}% bold math
\usepackage{color}
\usepackage{tabularx}
\usepackage{array}
\usepackage{amsmath}
\usepackage{stmaryrd}  %\llbracket \rrbracket

\begin{document}

\title{Strong coupling between propagating spin wave and microwave photons in a superconducting resonator}

\author{Yi Li}
\email{yili@anl.gov}
\affiliation{Materials Science Division, Argonne National Laboratory, Lemont, IL 60439, USA}

\author{Jinho Lim}
\affiliation{Department of Materials Science and Engineering, The Grainger College of Engineering, University of Illinois Urbana-Champaign, Urbana, IL 61801, USA}
\affiliation{Materials Research Laboratory, The Grainger College of Engineering, University of Illinois Urbana-Champaign, Urbana, IL 61801, USA}

\author{Xingzhi Wang}
\affiliation{Department of Materials Science and Engineering, The Grainger College of Engineering, University of Illinois Urbana-Champaign, Urbana, IL 61801, USA}
\affiliation{Materials Research Laboratory, The Grainger College of Engineering, University of Illinois Urbana-Champaign, Urbana, IL 61801, USA}

\author{Tomas Polakovic}
\affiliation{Physics Division, Argonne National Laboratory, Lemont, IL 60439, USA}

\author{Carissa Kiehl}
\affiliation{Materials Science Division, Argonne National Laboratory, Lemont, IL 60439, USA}
\affiliation{Physics and Astronomy Department, Carthage College, Kenosha, WI 53140, USA}

\author{Moojune Song}
\affiliation{Materials Science Division, Argonne National Laboratory, Lemont, IL 60439, USA}
\affiliation{Department of Physics, Korea Advanced Institute of Science and Technology,Daejeon 34141, Republic of Korea}

\author{Phuoc Cao Van}
\affiliation{Department of Materials Science and Engineering, Chungnam National University, Daejeon, 34134, Republic of Korea}

\author{Ralu Divan}
\affiliation{Center for Nanoscale Materials, Argonne National Laboratory, Argonne, IL 60439, USA}

\author{Ulrich Welp}
\affiliation{Materials Science Division, Argonne National Laboratory, Lemont, IL 60439, USA}

\author{Charudatta Phatak}
\affiliation{Materials Science Division, Argonne National Laboratory, Lemont, IL 60439, USA}

\author{Jong-Ryul Jeong}
\affiliation{Department of Materials Science and Engineering, Chungnam National University, Daejeon, 34134, Republic of Korea}

\author{Kab-Jin Kim}
\affiliation{Department of Physics, Korea Advanced Institute of Science and Technology,Daejeon 34141, Republic of Korea}

\author{Jian-Min Zuo}
\affiliation{Department of Materials Science and Engineering, The Grainger College of Engineering, University of Illinois Urbana-Champaign, Urbana, IL 61801, USA}
\affiliation{Materials Research Laboratory, The Grainger College of Engineering, University of Illinois Urbana-Champaign, Urbana, IL 61801, USA}

\author{Axel Hoffmann}
\email{axelh@illinois.edu}
\affiliation{Department of Materials Science and Engineering, The Grainger College of Engineering, University of Illinois Urbana-Champaign, Urbana, IL 61801, USA}
\affiliation{Materials Research Laboratory, The Grainger College of Engineering, University of Illinois Urbana-Champaign, Urbana, IL 61801, USA}

\author{Valentine Novosad}
\email{novosad@anl.gov}
\affiliation{Materials Science Division, Argonne National Laboratory, Lemont, IL 60439, USA}

\date{\today}

\begin{abstract}

We demonstrate strong coupling between propagating spin wave modes and microwave photons in superconducting resonator-magnetic thin film hybrid circuits. By fabricating the resonator directly on yttrium iron garnet thin films grown on rare-earth-free Y$_3$Sc$_2$Ga$_3$O$_{12}$ substrates, we achieve strong coupling of both Damon-Eshbach and backward-volume spin wave modes to the resonator, with coupling strengths exceeding both the magnon and photon damping rates. Furthermore, we observe nonreciprocal spin wave radiation of the hybrid magnonic mode in the Damon-Eshbach configuration, highlighting the potential for incorporating intrinsic spin-wave nonreciprocity into hybrid magnonic systems. These results open new avenues for integrating spin-wave magnonics with cavity magnonics, and for harnessing spin waves for potential applications in quantum information science.

\end{abstract}

\maketitle

\section{Introduction}

Hybrid quantum systems provide a versatile platform for coupling excitations from disparate physical systems and exploring their potential in quantum information processing \cite{KurizkiPNAS15,ClerkNPhys2020}. Hybrid magnonics, which cultivates coherent interaction of solid-state magnetic excitations or magnons \cite{LiJAP20,ZarePhysRep22}, offers unique material properties from magnetic systems, such as intrinsic magnon non-reciprocity and the capability of coupling to both microwave and light, for developing quantum information functionality such as on-chip isolators \cite{WangPRApplied21,LiIEDM22,OwensNPhys22} and microwave-to-optic transducers \cite{HisatomiPRB16,OsadaPRL16,ZhangPRL16,HaighPRL16,KusminskiyPRA16,KostylevJMMM19,ZhuOptica20}. In particular, the successful demonstration of magnon-qubit coupling \cite{TabuchiScience15,LachanceScience20,XuPRL23,RaniPRApplied25} unlocks the capability of single-magnon quantum manipulation to explore realistic quantum magnonic operations and magnon quantum sensing \cite{LachanceAPEx19,TricklePRL20,ChangarXiv25}.

%The architecture of those quantum magnonic systems usually includes a low-damping magnetic crystal, usually a single-crystal yttrium iron garnet (YIG) sphere, that strongly couples to a microwave cavity, which again coherently couples to a superconducting qubit as a single magnon detector. Thus,

Up to now, most experimental demonstrations of coherent magnon control have been focusing on Kittel modes, which are the uniform excitation of collective magnetic spins. The major material uniquenesses of the magnetic systems for quantum information functionality, however, rely on propagating spin waves. Examples include magnon nonreciprocity for on-chip isolators \cite{SchneiderPRB2008,JamaliSciRep2013,LiAPL23,JiangAPL23}, spin-wave-mediated spin qubit entanglement \cite{TrifunovicPRX13,AndrichnpjQI17,YuanPRA23,FukamiPNAS24}, and microwave-to-optical transduction with wavelength matching \cite{ZhuOptica20,ZhuPRApplied22,SekinePRApplied24}. Nevertheless, propagating spin waves are rarely explored experimentally in cavity magnonics \cite{BaiPRL15,ZhangJAP16}, especially at cryogenic temperature. Metallic ferromagnets suffer from conductivitylike damping enhancement at low temperatures \cite{Kambersky70,GilmorePRL07,KhodadadiPRL20}. The lowest-damping magnetic insulator for magnonics, yttrium iron garnet (YIG) films, are usually grown on Gd$_3$Ga$_5$O$_{12}$ (GGG) substrates, but GGG substrates suffer from their complex magnetic order at cryogenic temperature and strong microwave absorption \cite{LiuPRB18,KosenAPLMater19,SerhanpjSpintronics24}. Thus, alternative substrates compatible with cryogenic microwave engineering is highly desired for low-damping YIG thin film growth and spin-wave-based cavity magnonics at low temperature.

%Moreover, it is challenging to find a low-damping magnetic thin film system at cryogenic temperatures. Metallic ferromagnets usually exhibit conductivity-like damping mechanism and the magnetic damping will significantly increase when the temperature decrease. For YIG thin films which exhibit the lowest magnetic damping at room temperature, the cryogenic microwave performance is strongly undermined by the spin texture of the Gd3Ga5O12 substrates typically used for epitaixal YIG growth.

In this work, we develop superconducting hybrid magnonic circuits which achieve strong coupling between propagating spin wave and microwave photons. In particular, we fabricate superconducting coplanar resonators on YIG thin films grown on Y$_3$Sc$_2$Ga$_3$O$_{12}$ (YSGG) substrates. The use of rare-earth-free YSGG substrates addresses the challenge of integrating YIG thin films with cryogenic superconducting circuits. We employ a quarter-wavelength coplanar waveguide superconducting resonator design with micrometer-scale lateral widths matched to the short-wavelength spin wave modes and millimeter-scale longitudinal lengths matched to the microwave photon wavelength, enabling coherent coupling between magnons and photons with drastically different wavelengths. In the Damon-Eshbach (DE) or backward volume (BV) spin wave configuration, we show that the spin wave mode coupled to the resonator exhibits a positive or negative frequency offset relative to the quasi-Kittel mode coupled to the microwave feed line, consistent with the respective spin wave dispersions. Furthermore, by modifying the microwave coupler that detects the spin wave mode excited by the superconducting resonator, we demonstrate nonreciprocal detection of the hybrid magnonic mode, attributed to the intrinsic nonreciprocity of DE spin wave mode. Our results open new opportunities in building cryogenic hybrid magnonic circuits with propagating spin waves for exploring quantum magnonics.

\begin{figure}[htb]
 \centering
 \includegraphics[width=3.0 in]{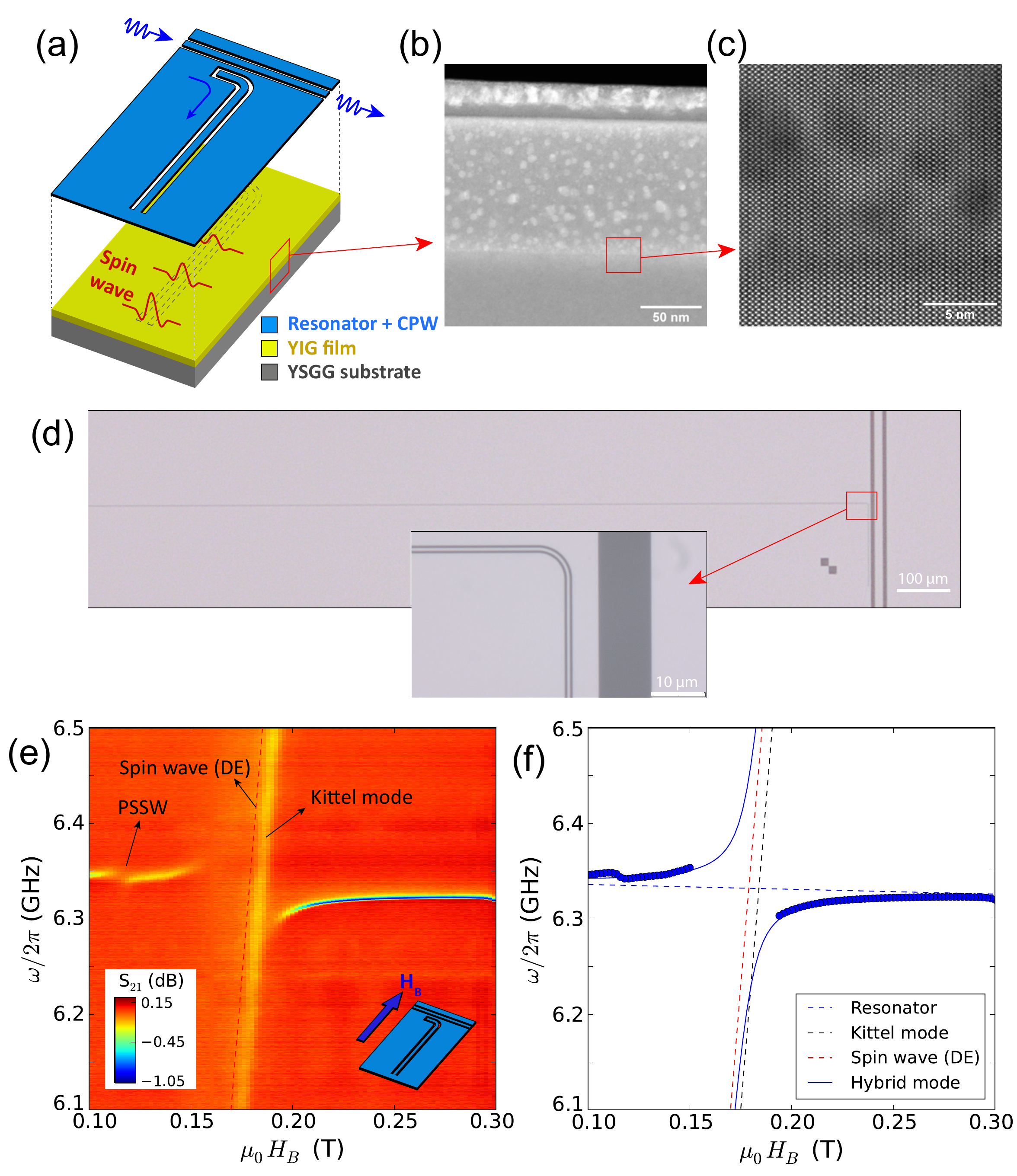}
 \caption{(a) Illustration of the YIG-$\lambda/4$ resonator design. (b-c) TEM image of the YIG/YSGG thin film. (d) Optical microscope image of a $\lambda/4$ resonator on a YSGG substrate with a signal linewidth $w=2$ $\mu$m. (e) Mode anticrossing between the DE spin wave mode and the resonator mode along with the continuous Kittel mode for the $\lambda/4$ resonator with $w=2$ $\mu$m. (f) Extracted peak positions and the fitting curves.}
 \label{fig1}
\end{figure}

The configuration of the superconducting hybrid magnonic systems are illustrated in Fig. \ref{fig1}(a), where superconducting resonators were fabricated from NbN superconducting thin films that is grown on YIG films. The YIG films were deposited by sputtering onto YSGG(111) substrates. The films were subsequently annealed in air at 850 $^\circ$C for 3 hours, followed by slow cooling to room temperature at a rate of 30 $^\circ$C/hour. The epitaxial quality of the YIG films was confirmed by transmission electron microscopy (TEM) [Fig. \ref{fig1}(b-c)]. The dark circular features correspond to strain-induced local lattice rotations and will be discussed elsewhere. For the superconducting layer, we first deposited a SiN(10 nm) layer on top of the YIG to minimize the magnetic proximity effect \cite{BuzdinRMP05}, and then deposited a NbN(30 nm) layer for resonator fabrication. The $\lambda/4$ resonator and the microwave feedline [Fig. \ref{fig1}(d)] were patterned using electron-beam lithography, and etched using reactive ion etching.

\begin{figure}[htb]
 \centering
 \includegraphics[width=3.0 in]{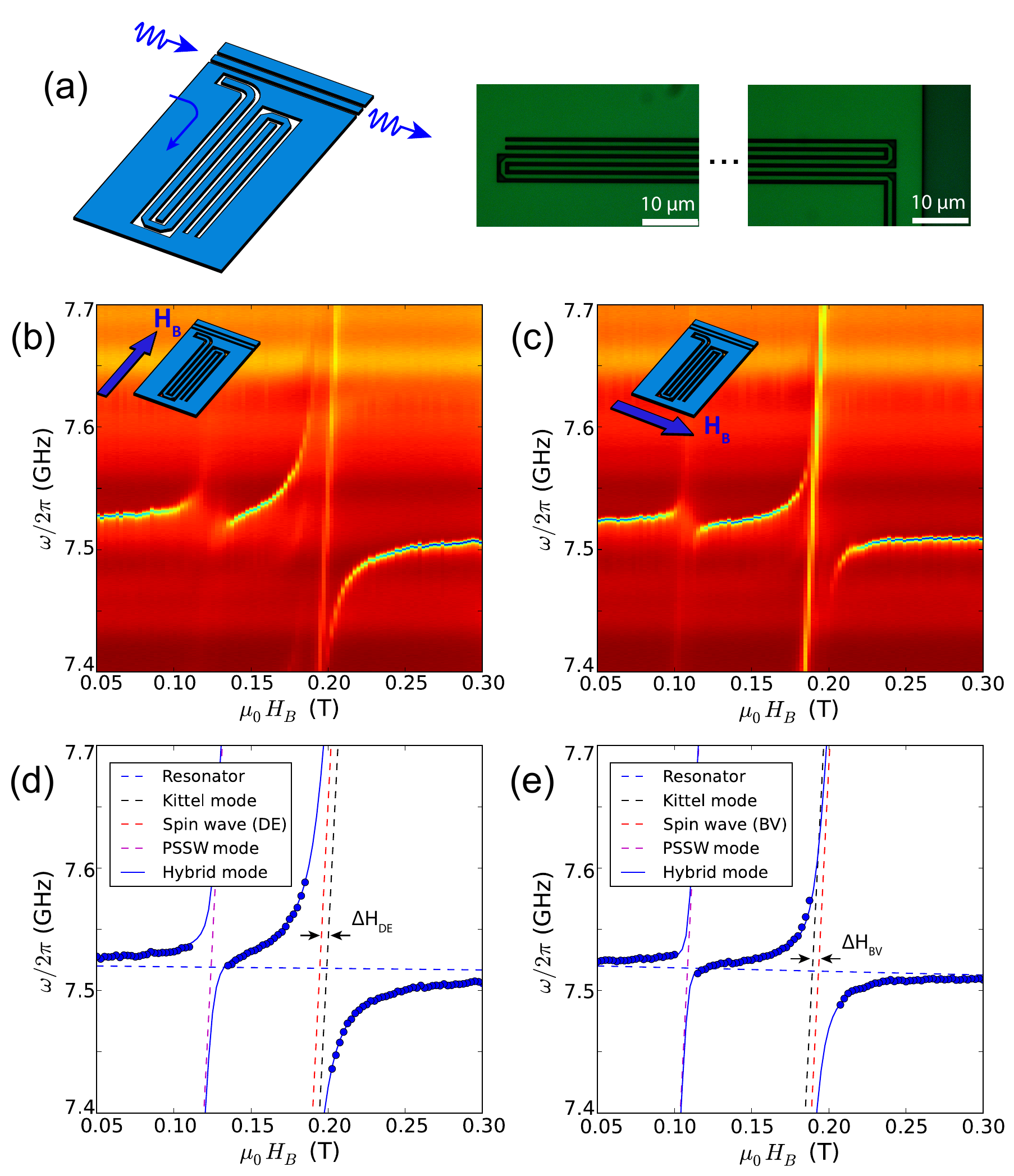}
 \caption{(a) Schematic and optical microscope of meander superconducting reonator design with $w=1$ $\mu$m. (b-c) VNA spectra of the meander resonator with in-plane magnetic field (b) parallel and (c) perpendicular to the longitudinal direction of the resonator. (d-e) Extracted peak positions and the fitting curves.}
 \label{fig2}
\end{figure}

Two types of coplanar waveguide resonators have been designed for coupling to spin wave. The first type contains a straight quarter wavelength \textcolor{black}{[$\lambda/4$, Fig. \ref{fig1}(d)]} resonator. The second type contains a meander design \textcolor{black}{[meander, Fig. \ref{fig2}(a)]}, which is equivalent to the $\lambda/4$ resonator but incorporates three folds for better wavelength selectivity \textcolor{black}{\cite{Vlaminck08,NeusserPRL10,LiuNComm2018}}. To test the spin wave frequency dispersion versus wavenumber, we fabricated resonators with different signal line widths, i.e. $w = 2$ $\mu$m and 0.5 $\mu$m for the $\lambda/4$ resonator and $w = 1$ $\mu$m for the meander resonator; see the Supplemental Information for more details in resonator dimensions and the matching wavelengths \cite{supplement}. The total lengths of the resonators for both designs are 3 mm, but due to different impedance environments, their frequencies are slightly different. The YIG films used in the devices were grown at Argonne for the $\lambda/4$ resonators and at KAIST for the meander resonators. The KAIST YIG films exhibit better lattice match to the YSGG substrate and lower magnetic damping at cryogenic temperatures.

\textcolor{black}{\textit{Strong coupling with $\lambda/4$ resonator}}. Evidence of strong coupling between spin wave and microwave photons is shown in the  vector network analyzer (VNA) spectra of the $w=2$ $\mu$m device in Fig. \ref{fig1}(e). When an in-plane magnetic field is applied along the longitudinal direction of the $\lambda/4$ resonator [see Fig. \ref{fig1}(e) inset], an avoided crossing emerges in the resonator mode, accompanied by a continuous mode traversing the anticrossing gap. The avoided crossing indicates coherent coupling of the resonator to the DE spin wave mode, whose wavevector $\vec{k}$ is perpendicular to both the signal line and the magnetic field ($\vec{k}\perp\vec{H}_B$). The continuous mode corresponds to the quasi-Kittel mode of the YIG film coupled to the wide microwave feedline, which has a signal line width of 20 $\mu$m, much larger than $w$. From the spectra we extract a magnon damping rate of $\kappa_m/2\pi=70$ MHz for the YIG/YSGG film at low temperature. Fig. \ref{fig1}(f) shows fits to the magnon-photon hybrid modes, $\omega_\pm=(\omega_c+\omega_m)/2\pm\sqrt{(\omega_c-\omega_m)^2/4+g^2}$ where $\omega_c$, $\omega_m$, $g$ denote the resonator frequency, DE spin wave mode frequency, and the magnon-photon coupling strength, respectively. The extracted spin wave mode (red dashed line) deviates from the Kittel mode (black dashed line) towards lower magnetic fields or equivalently higher frequencies, consistent with the DE spin-wave mode dispersion. We obtain a spin-wave-photon coupling strength of $g/2\pi=115$ MHz. Same feature is observed for the 0.5~$\mu$m resonator; see the Supplemental Materials for additional details \cite{supplement}

%Moreover, because the chirality of the microwave field from the CPW resonator only matches spin waves propagating in one direction \cite{LiAPL23}, the hybrid magnonic mode radiates spin waves unidirectionally along $\vec{H}_B\times \hat{n}$, where $\hat{n}$ is the film normal from the YIG towards the CPW resonator [Fig. \ref{fig2}(b)].

%To further reduce the magnon damping, we employ improved YIG(100 nm)/YSGG films grown at KAIST, which exhibit better lattice match to the YSGG substrate and lower magnetic damping at cryogenic temperatures \cite{supplement}. In addition, we adopt a meander resonator design, which is equivalent to a $\lambda/4$ resonator but incorporates three folds, as shown in Fig. \ref{fig2}(a). This geometry enhances wavelength selectivity for the spin-wave mode, thereby reducing magnetic relaxation associated with frequency dispersion. The signal-line width of the meander resonator is designed to be $w=1$ $\mu$m, enabling coupling to spin waves with different wavenumbers.

\textcolor{black}{\textit{Strong coupling with meander resonator.}} In Fig. \ref{fig2}(b) and (c) we show the measured VNA spectra of the meander resonator for the DE and BV magnetic field configurations. Compared with the measured quasi-Kittel mode, the spin wave mode coupled to the meander resonator exhibits a negative field offset in the DE configuration and a positive field offset in the BV configuration. Note that the field offset is larger than in Fig. \ref{fig1}(e) because the signal line width is smaller and the excited spin wave modes have larger wavevectors.

From the quasi-Kittel mode, we extract an improved magnon damping rate of $\kappa_m/2\pi=34$ MHz for the new YIG/YSGG film. In addition, we observe an extra mode anticrossing at lower fields in both configurations, corresponding to the perpendicular standing spin wave (PSSW) mode \cite{LiPRL16}. The coupling strength to this mode is clearly stronger than observed for the straight $\lambda/4$ resonator in Fig. \ref{fig1}(e). This enhancement originates from the meander geometry, which increases the microwave field inhomogeneity along the YIG film-thickness direction and thereby improves coupling to the PSSW mode. However, if we calculate the effective exchange field using $\mu_0H_{ex}=(2A_{ex}/M_s)k_\perp^2$ with the YIG exchange constant $A_{ex}=2.6$ pJ/m and magnetization $\mu_0M_s=0.175$ T, and $k_\perp=\pi/t$ where $t=100$ nm, we obtain $\mu_0H_{ex}=37$ mT, which is significantly smaller than the experimentally observed value of 80 mT. One possibility is that the YIG/YSGG interface has a strong surface pinning due to the lattice mismatch, which leads to a different PSSW boundary condition as $k_{\perp,n}=(n+1/2)\pi/t$. The new exchange field is then calculated as $\mu_0H_{ex}=(2A_{ex}/M_s)(k_{\perp,1}^2-k_{\perp,0}^2))=74$ mT, agreeing more closely with the experimental value.

\begin{figure}[htb]
 \centering
 \includegraphics[width=3.0 in]{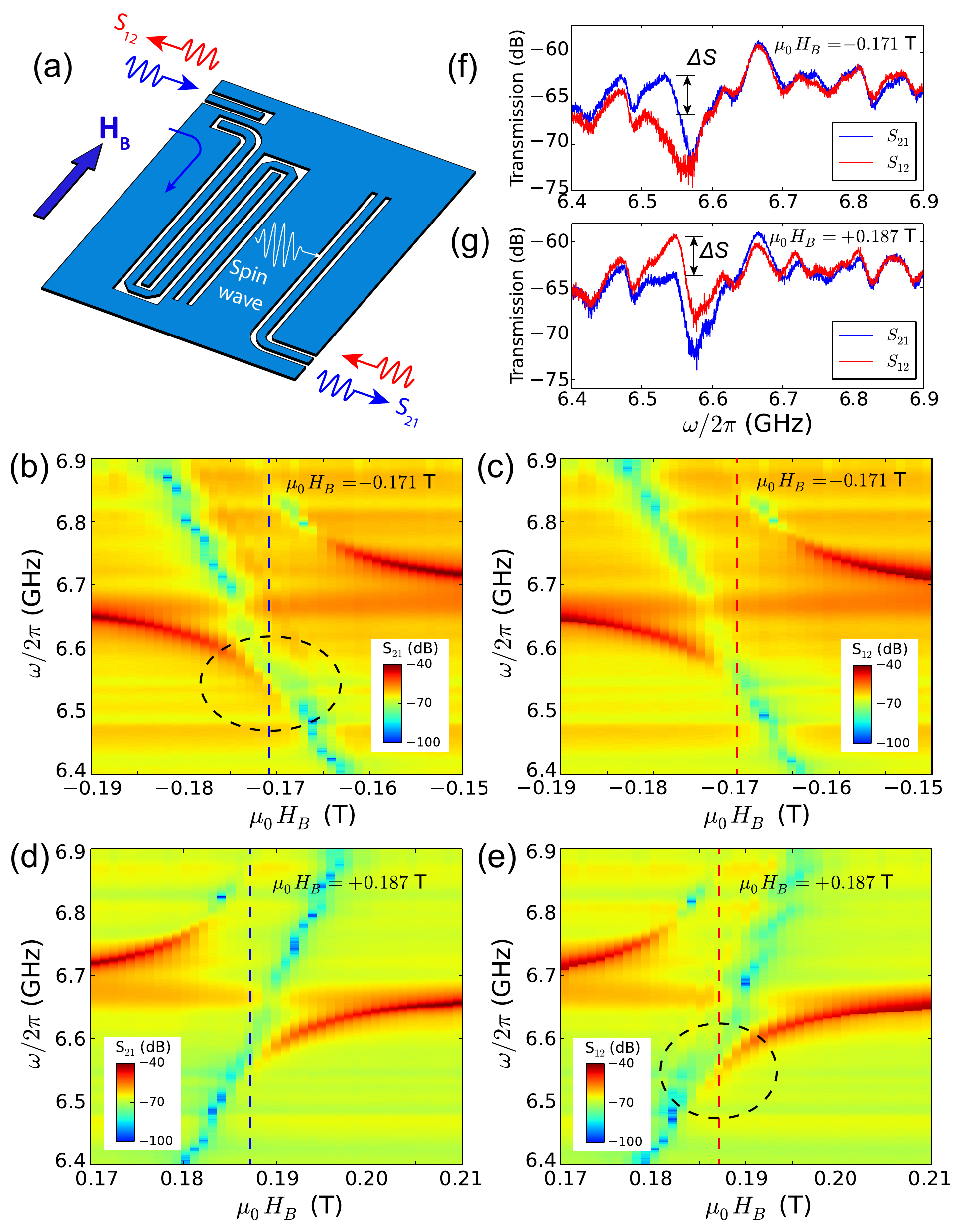}
 \caption{(a) Schematic of the resonator design with parallel CPW detecting port and magnetic field parallel to the resonator and detecting port. (b-c) VNA spectra of (b) $S_{21}$ and (c) $S_{12}$ with negative field. (d-e) VNA spectra of (d) $S_{21}$ and (e) $S_{12}$ with positive field. (f-g) Lineshape comparison at (f) $\mu_0H_B=-0.171$ T and (g) $+0.187$ T. The field locations are labelled in (b-e).}
 \label{fig4}
\end{figure}

\textcolor{black}{\textit{Nonreciprocity with meander resonator.}} To further demonstrate the spin wave character of the hybrid magnonic mode, we modify the coupler design by fabricating a parallel CPW adjacent to the meander resonator as the detection port, as shown in Fig. \ref{fig4}(a). \textcolor{black}{The CPW coupler has the same signal-line width and signal-ground gap as the $\lambda/4$ resonator, enabling efficiently coupling to the spin-wave mode radiated from the resonator.} \textcolor{black}{The CPW couple-to-resonator distance is 5 $\mu$m, smaller than the reported spin wave propagation distance of 8.7 $\mu$m measured in YIG/GGG with the same thickness and wavelength \cite{LiAPL23}, which balances the spin wave propagation loss with the capacitive crosstalk.} Fig. \ref{fig4}(b) and (c) show the VNA transmission spectra with the microwave excitation applied to the excitation port [$S_{21}$, Fig. \ref{fig4}(b)] and to the detection port [$S_{12}$, Fig. \ref{fig4}(c)], respectively. Both measurements were performed during the same magnetic-field sweep. In the mode hybridization regime of the lower branch [circle dash in Fig. \ref{fig4}(b)], $S_{21}$ exhibits stronger transmission than $S_{12}$. Upon reversing the magnetic field, the nonreciprocity is also reversed, and the stronger transmission branch appears in $S_{12}$ [circle dash in Fig. \ref{fig4}(e)]. A comparion of the lineshapes is shown in Fig. \ref{fig4}(f) at $\mu_0H_B=-0.171$ T and in Fig. \ref{fig4}(g) at $\mu_0H_B=+0.187$ T, where the magnon and photon modes are frequency degenerate. The field difference is due to hysteresis in the superconducting magnet. A power isolation of $|S_{21}-S_{12}|=4$ dB is measured at the hybrid-mode transmission peak for both field polarities.

The observed nonreciprocity confirms the role of the Damon-Eshbach spin wave in the hybrid magnonic mode. \textcolor{black}{In our control experiments as shown in the Supplemental Materials \cite{supplement}, this nonreciprocity is absent when the magnetic field is rotated by 90$^\circ$ to the backward-volume configuration. We also show that using the same CPW antenna geometry as in Fig. \ref{fig4}(a), a broad-band spin wave nonreciprocity of 10 dB can be measured under the Damon-Eshbach configuration. The results show that the intrinsic spin wave nonreciprocity can be indeed integrated into hybrid magnonics, providing realistic potential for on-chip integration of microwave isolators which is highly desired in quantum information science. \cite{JiangAPL23}}

\begin{figure}[htb]
 \centering
 \includegraphics[width=3.0 in]{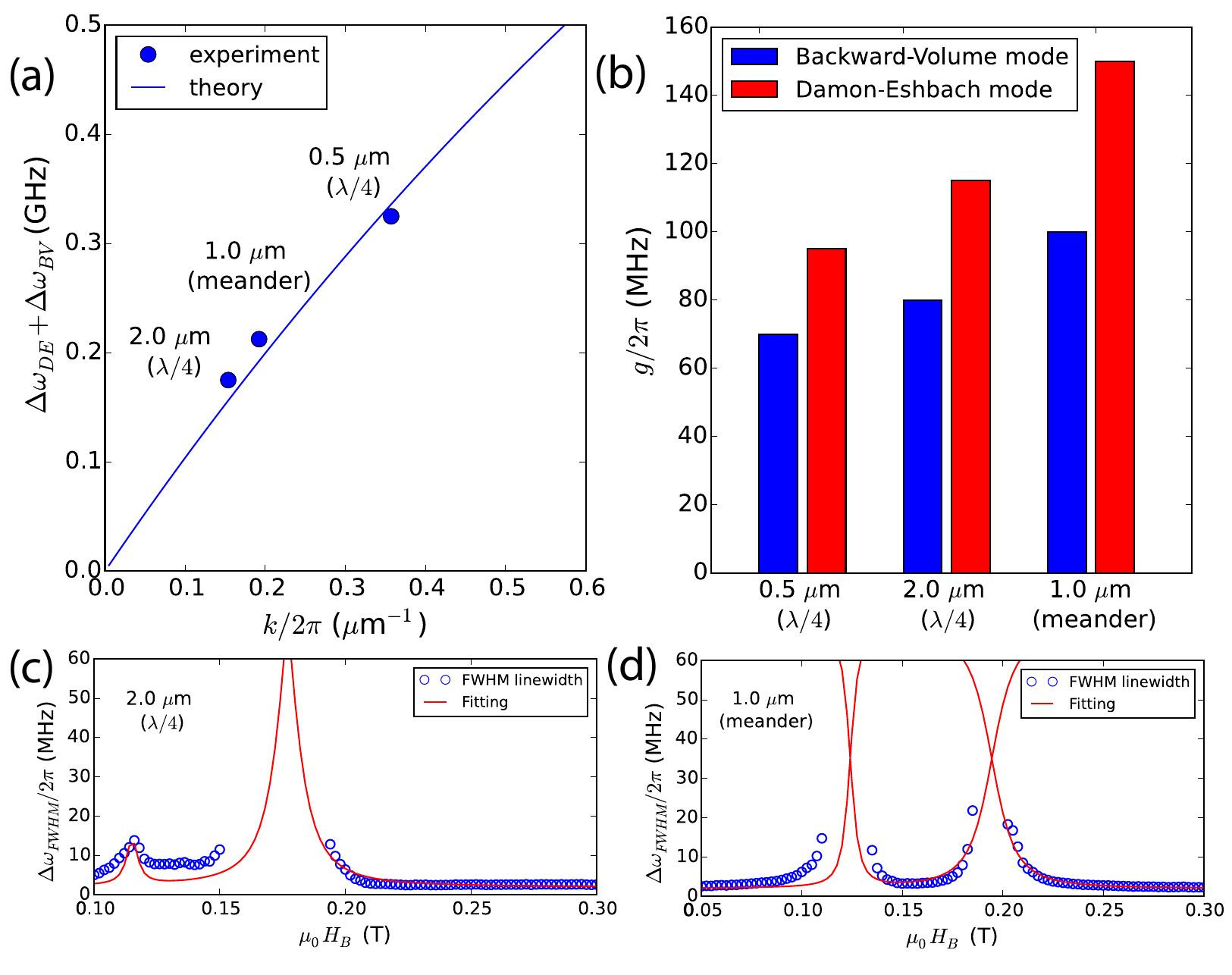}
 \caption{(a) Extracted frequency offset between the Damon-Eshbach and backward-volume spin wave modes for three difference resonator devices. The blue curve is the theoretical prediction. (b) Extracted magnon-photon coupling strength for the three devices with two different $H_B$ configuration. Damon-Eshbach mode: $H_B\perp k$; backward-volume mode: $H_B\| k$. (c) Extracted FWHM linewidth of the 2-$\mu$m $\lambda/4$ resonator from Fig. \ref{fig2}(c). (d) Extracted FWHM linewidth of the 1-$\mu$m meander resonator from Fig. \ref{fig3}(b). Both are measured under the DE mode configuration. Red curves are fits using Eq. (\ref{eq3}). Used fit parameters are: $g/2\pi=115$ MHz in (a) and 150 MHz in (b).}
 \label{fig3}
\end{figure}

\textcolor{black}{\textit{Analysis of the extracted fit parameters.}} In Fig. \ref{fig3}(a) we plot the sum of frequency offset, $\Delta\omega_\text{DE}-\Delta\omega_\text{BV}$, as a function of wavenumber $k$ for the three devices (two $\lambda/4$ resonators with $w=0.5$ and 2 $\mu$m and one meander resonator with $w=1$ $\mu$m). The experimental values of $\Delta\omega_\text{DE}$ and $\Delta\omega_\text{BV}$ are calculated from the field offset, $\Delta H_\text{DE}$ and $\Delta H_\text{BV}$ as shown in Fig. \ref{fig2} (d) and (e), times the slope of the $\omega-H$ dispersion. Note that $\Delta\omega_\text{DE}, \Delta H_\text{DE} >0$ and $\Delta\omega_\text{BV}, \Delta H_\text{BV} <0$. The experimental $k$ values are calculated from the geometry of the resonator design with different signal line widths. We obtain the wavelengths with maximum coupling efficiency as $\lambda=2.8$, 5.2, and 6.5 $\mu$m for $w=0.5$ ($\lambda/2$), 1.0 (meander) and 2.0 ($\lambda/2$) $\mu$m, respectively; see the Supplemental Materials for details \cite{supplement}. For the theoretical curve of $\Delta\omega_\text{DE}-\Delta\omega_\text{BV}$, we take:
\begin{align}
    \omega_\text{DE}(k) &= \gamma\mu_0\sqrt{H(H+M_s)+{M_s^2\over 4}(1-e^{-2kt})}\\
    \omega_\text{BV}(k) &= \gamma\mu_0\sqrt{H(H+M_s \times{1-e^{-kt} \over kt})}
\end{align}
with $\omega_0=\gamma\mu_0\sqrt{H(H+M_s)}$ as the Kittel mode and $\Delta\omega_\text{DE(BV)}=\omega_\text{DE(BV)}-\omega_0$. We use the saturation magnetization $\mu_0M_s=0.175$ T for YIG, the gyromagnetic ration $\gamma/2\pi=28$ GHz/T, and the thickness $t=100$ nm for the theoretical curve. The excellent agreement between the experiments and theory confirms the role of the spatially nonuniform spin wave modes which strongly couple with the microwave photon mode of the resonator.

%In the DE configuration, the 1-$\mu$m meander design shows a coupling strength of 150 MHz, which is 58\% larger than the 0.5-$\mu$m $\lambda/4$ resonator (95 MHz) and 30\% larger than the 2-$\mu$m $\lambda/4$ resonator (115 MHz).

Fig. \ref{fig3}(b) summarizes the extracted magnon-photon coupling strength for the three devices. The DE configuration exhibits a coupling strength that is 43\% larger than that of the BV configuration on average. This difference arises because in the DE configuration, both the in-plane and vertical components of the microwave field generated from the resonators are orthogonal to the YIG magnetization and can contribute to the coupling. In contract, for the BV configuration, only the vertical component contributes to the coupling, resulting in weaker spin-wave coupling which is consistent with previous observations in CPW resonator systems\cite{LiPRR23}. We also emphasize the enhanced coupling strength achieved with the meander resonator design compared to the straight CPW design. This enhancement primarily originates from improved mode overlap between the spatially alternating microwave field of the meander geometry and the sinusoidal spin wave mode. Notably, varying the lateral dimensions of the resonator does not necessarily change the coupling strength, since it is predominantly determined by the mode-volume ratio between the magnon ($V_m$) and resonator ($V_c$) systems, with $g\sim \sqrt{V_m/V_c}$ \cite{LiJAP20}. In this ratio, the lateral area cancels out, leading to only a weak dependence on resonator width.

Fig. \ref{fig3}(c) and (d) show the extracted full-width-half-maximum (FWHM) linewidths of the hybrid magnonic modes for the 2-$\mu$m $\lambda/4$ resonator [Fig. \ref{fig1}(e)] and the 1-$\mu$m meander resonator [Fig. \ref{fig2}(b)] under the DE configuration. The red curves show the theoretical calculations of the linewidth evolution based on the analytical equation\cite{supplement}:
\begin{equation}
    \kappa_\pm={\kappa_m \over 2}({\pm\Delta\omega/2 \over \sqrt{{\Delta\omega^2 / 4}+g^2}}+1)+{\kappa_c \over 2}({\mp\Delta\omega/2 \over \sqrt{{\Delta\omega^2 / 4}+g^2}}+1)
    \label{eq3}
\end{equation}
where $\Delta\omega = \omega_c-\omega_m$ denotes the frequency detuning. We use the magnon damping $\kappa_m$ previously extracted from the quasi-Kittel mode, to represent the intrinsic spin wave damping [70 MHz for Fig. \ref{fig3}(c) and 34 MHz for Fig. \ref{fig3}(d)]. The photon damping rate is determined to be $\kappa_c/2\pi=1.1$ MHz for both devices when the magnon mode is far detuned from the resonator frequency. For the PSSW mode in the 2-$\mu$m $\lambda/4$ resonator, the magnon-photon coupling strength is smaller than the magnon damping. In this regime, we fit the date using the Purcell-limit damping rate expression \cite{XiongSciRep20,ZollitschNComm23}; see the Supplemental Materials for details \cite{supplement}.

For the 1-$\mu$m meander resonator, the theoretical curve reproduces the experiments well, supporting our assumption of taking the magnon damping of the quasi-Kittel modes adequately represents the propagating spin-wave damping. Our results therefore demonstrate strong magnon-photon coupling in the device, with a cooperativity of $C=g^2/\kappa_c\kappa_m=662$ for the DE spin wave mode ($g/2\pi=150$ MHz) and $C=294$ for the BV spin wave mode ($g/2\pi=100$ MHz). In both cases, the coupling strength satisfies $g>\kappa_m,\kappa_c$, confirming the strong-coupling regime. For the $\lambda/4$ resonator, the experimentally measured linewidth exceeds the theoretical prediction in the lower-field branch ($\mu_0H_B\sim 0.15$ T). This discrepancy suggests that the spin-wave damping is larger than the Kittel-mode magnon damping, rendering it uncertain whether the condition $g>\kappa_m$ is strictly satisfied.

We estimate the spin-wave radiation damping rate based on the spin wave propagation away from the resonator antennas. The damping rate can be calculated as $\kappa_{sw}=1/ t_{sw}$ where $t_{sw}=l/v_g$ is the escape time, $l$ is the effective antenna width, and $v_g$ is the spin-wave group velocity. For the 0.5-$\mu$m, 2-$\mu$m $\lambda/4$ resonators and the 1-$\mu$m meander resonator, we take effective antenna widths of $l=2$ $\mu$m, 6 $\mu$m and 10 $\mu$m, and the calculated group velocities as $v_g=333$ m/s, 435 m/s, and 413 m/s, respectively; see the Supplemental Materials for details \cite{supplement}. From these values, we obtain $\kappa_{sw}/2\pi=26.5$ MHz for the 0.5-$\mu$m $\lambda/4$ resonators, 11.5 MHz for the 2-$\mu$m $\lambda/4$ resonators, and 6.6 MHz for the 1-$\mu$m meander resonator. These values are smaller than $\kappa_m$, meaning that the spin wave damping is dominated by the intrinsic magnon relaxation within the YIG film.

%The additional spin-wave damping observed in the $\lambda/4$ resonator, compared with the meander resonator, may arise from two contributions. First, the $\lambda/4$ resonator antennas couples to a broader spin-wave band due to its fewer spatial repeats compared to the meander design, resulting in effective linewidth broadening. Second, the excited spin waves propagate away from the antennas, leading to additional energy dissipation. In the meander resonator, the larger effective antenna width increases the time required for the excited spin waves to escape the antenna region, thereby reducing this propagation-induced damping.

Our experiments show that even without lateral spin wave confinement, it is possible to achieve strong coupling with microwave photons. This is mainly due to the slow group velocity of spin wave, leading to $\kappa_{sw}\ll g$. For the potential of using propagating spin waves for distant resonator coupling, one needs to deliberately increase the weight of $\kappa_{sw}$ over magnon damping $\kappa_m$ to emit spin wave before it decays. Recent reports show $\kappa_m/2\pi$ below 10 MHz at cryogenic temperatures for YIG thin films grown on cryogenic-compatible substrates \cite{ChumakarXiv25, ChumakarXiv25_2, YoussefarXiv25, GambardellaarXiv25}, showing promise for building magnonic circuits that incorporate coherent magnon-photon interconversion.

The use of coplanar superconducting circuits and YIG thin films enables versatile spin wave control and engineering. For example, $\kappa_{sw}$ can be further reduced by increasing $l$, linewidth broadening due to spin wave bandwidth can be suppressed by increasing the meander layers, and $v_g$ can be controlled by changing the thickness of the YIG film. The architecture on-chip hybrid magnonic circuit can be also extended to the quantum regime by integrating magnetic-field-compatible qubits, such as semiconductor spin qubits \cite{BurkardRMP23} and single-electron qubits \cite{KoolstraNComm19,ZhouNature22}, as single-propagating-magnon detectors.

In conclusion, we experimentally demonstrate strong coupling between propagating spin wave and microwave photons in on-chip superconducting resonator circuits integrated on YIG/YSGG thin films. Our results extend cavity magnonics to propagating spin waves at cryogenic temperatures, which may open new pathways towards quantum information processing with spin waves beyond Kittel modes.

\section{Acknowledgement}

All research work, including sample preparation, microwave characterization, and data analysis, were primarily supported by the U.S. DOE, Office of Science, Basic Energy Sciences, Materials Sciences and Engineering Division under contract No. DE-SC0022060. The lithographic patterning and fabrication of the superconducting resonators performed at the Center for Nanoscale Materials, a U.S. Department of Energy Office of Science User Facility, was supported by the U.S. DOE, Office of Basic Energy Sciences, under Contract No. DE-AC02-06CH11357.

%\bibliography{ref_propagating_YIG_magnon}

%apsrev4-2.bst 2019-01-14 (MD) hand-edited version of apsrev4-1.bst
%Control: key (0)
%Control: author (8) initials jnrlst
%Control: editor formatted (1) identically to author
%Control: production of article title (0) allowed
%Control: page (0) single
%Control: year (1) truncated
%Control: production of eprint (0) enabled
%

\end{document}